# Aharonov-Bohm electron interferometer in the integer quantum Hall regime

F. E. Camino, W. Zhou, and V. J. Goldman
*Department of Physics and Astronomy, Stony Brook University, Stony Brook, New York 11794-3800, USA*

We report experiments on a quantum electron interferometer fabricated from high mobility, low density AlGaAs/GaAs heterostructure material. In this device, a nearly circular electron island is defined by four front gates deposited in etched trenches. The island is separated from the 2D electron bulk by two nearly open constrictions. In the quantum Hall regime, two counterpropagating edge channels are coupled by tunneling in the constrictions, thus forming a closed electron interference path. For several fixed front gate voltages, we observe periodic Aharonov-Bohm interference oscillations in four-terminal resistance as a function of the enclosed flux. The oscillation period $\Delta B$ gives the area of the interference path $S$ via the quantization condition $S = h/e\Delta B$. We experimentally determine the dependence of $S$ on the front gate voltage, and find that the Aharonov-Bohm quantization condition does not require significant corrections due to the confining potential. These results can be interpreted as a constant integrated compressibility of the island with respect to the front gates. We also analyze experimental results using two classical electrostatics models: one modeling the 2D electron density due to depletion from an etch trench, and another modeling the gate voltage dependence of the electron density profile in the island.

## I. INTRODUCTION

Quantum interference of 2D electrons around a quantum antidot subjected to a quantizing magnetic field has been used experimentally to determine the fractional charge of Laughlin quasiparticles of the surrounding quantum Hall condensate.[1,2] Recent experiments on devices in the inverse geometry, where quantized electron paths circle a 2D electron island, have reported observation of an Aharonov-Bohm "superperiod", implying fractional statistics of Laughlin quasiparticles.[3] The layout of the present interferometer looks qualitatively similar to a "Coulomb Island",[4,5] but the electron island is larger. The principal difference is that the constrictions are nearly open, so that no Coulomb blockade or conductance steps are observed at zero magnetic field. In the integer quantum Hall (QH) regime, the Landau level filling in the constrictions is nearly equal to that in the 2D bulk.

In this paper we report electron quantum interference experiments in the integer QH regime performed with an electron interferometer device, Fig. 1. In this device, counterpropagating edge channels[6-9] enclose a lithographically defined 2D electron island, and tunneling in the two nearly open, tunable constrictions completes the electron path, thus allowing an Aharonov-Bohm-like interference regime. When tunneling between the edge channels occurs, in the quantum-coherent regime, Aharonov-Bohm oscillations with period $\Delta B$ are expected in the four-terminal resistance $R_{XX} = V_X/I_X$ as a function of the magnetic field $B$. In the quantum limit, each oscillation signals the alignment of a quantized electron state encircling the 2D electron island with the chemical potential $\mu$.

In each spin-polarized Landau level (LL), the single-electron states are quantized by the Aharonov-Bohm condition: the magnetic flux $\Phi$ through the area of an encircling orbital $S_m$ satisfies $\Phi = BS_m = m\Phi_0$, where $m$ is the quantum number of the orbital and $\Phi_0 = h/e$ is the fundamental flux quantum.[10,11] Thus, $S_m = mh/eB = 2\pi m\ell^2$, where $\ell = \sqrt{\hbar/eB}$ is the magnetic length, and the area for each electron state per spin-polarized LL is $S_{m+1} - S_m = 2\pi\ell^2 = h/eB$. These quantization conditions apply as well to an interacting 2D electron fluid with microscopically uniform density, so long as no phase transition to a charge density wave (such as a striped or "bubble") ground state occurs.

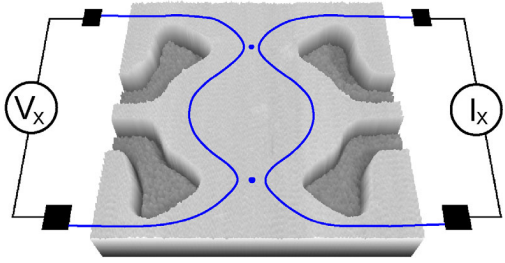

**FIG. 1.** An atomic force microscope (AFM) image of the central region of an electron interferometer device. A nearly circular region of lithographic radius $R = 1300$ nm was defined in AlGaAs/GaAs heterojunction material by chemically-etched trenches. Au/Ti metallization deposited in the etch trenches forms the front gates of the device. The constrictions are 1200 nm wide. Four Ohmic contacts (black squares) are in fact positioned at the corners of a $\sim 4 \times 4$ mm sample. On a quantum Hall plateau, with a quantizing magnetic field normal to the 2D electron plane, current flows along the counterpropagating edge channels (blue lines). Tunneling (represented by dots) occurs in the two wide constrictions, when the edge channels are close enough, thus allowing the electrons to perform a closed path around the 2DES island. The Aharonov-Bohm interference signal is detected as oscillations in $R_{XX} \equiv V_X / I_X$.

The 2D electron island in this sample is large, containing ~2000 electrons. Thus the electron density profile is expected to be determined mostly by the classical electrostatics of neutralizing the positively charged donors and the electric field of the gates, if biased. The island basic confinement is produced by the etch trenches which remove the donors. GaAs is known to have the "surface Fermi level pinning", due to a large density of mid-gap surface electron states, which has been successfully modeled by self-consistent depletion of donors, including a negative surface charge density. The surface depletion results in the 2D electron density being less than the donor density (because some donor electrons go to the surface), and, important for the present samples, an additional etched mesa sidewall depletion due to the free surface of the etch trenches.

The main effect of the external confinement potential $U(r)$ is to lift the massive degeneracy of the single-electron states in each Landau level. In the first order perturbation theory, the quantization $S_m = 2\pi m \ell^2$ is not affected by the confinement [$U(r)$ is simply added to the cyclotron and spin energies]. For the QH filling[12] $i = 1$ (the only QH filling presented in this paper), each $R_{XX}$ oscillation corresponds to a change by one in the number of electron states within $S_\mu$, the area enclosed by the path of the electron at energy $\mu$.[2,11] Thus, as a function of $B$, when $S_\mu$ is nearly fixed by the confining potential, the flux period $\Delta\Phi$ is one $\Phi_0$ for each oscillation, similar to the quantum antidots,[1,11] so that $S_\mu = \Phi_0 / \Delta B$.

We stress that changing $B$ does not change appreciably the number of electrons in the large island (this would lead to an enormous Coulomb energy). Instead, changing $B$ changes the density of states in each Landau level, so that the same number of island electrons occupies the same total number of states, but their distribution between various Landau levels changes: the Landau level filling $\nu = nh/eB$ changes while density $n$ is fixed. On the $i = 1$ QH plateau, in the 2D bulk $\mu$ is in the localized states between the lowest and the next spin-split Landau levels. Chemical potential in the island is determined by the $\mu$ in the bulk, our results (presented in this paper) show that apparently the radial position of the $i = 1$ edge channel circling the island is nearly fixed by the confining potential, that is, $S_\mu$ is nearly constant. The quantum number $m_\mu$ of the state $S_m$ corresponding to $S_\mu$ regularly changes in steps of one, $m_\mu \propto B$ since the area per state $2\pi \ell^2 \propto 1/B$. If the occupation of the states in the electron island were a step-function, 1 for $m \leq m_\mu$ and 0 for $m > m_\mu$, then $S_\mu$ must change, and the number of electrons within $S_\mu$ must change. Instead, as evidenced by the experiments reported here, both $S_\mu$ and the number of electrons within $S_\mu$ are nearly constant. What happens is that the electron occupation is not described by the step-function. Increasing $B$ increases density of states in a LL, thus accommodating the same number of electrons is accompanied by creation of unoccupied states (that is, holes) in the otherwise filled LL. Likewise, decreasing $B$ is accompanied by filling of the states in the next, otherwise empty LL. Such Landau level quasiparticle or quasihole creation allows to accommodate a fixed number of island electrons (as dictated by the self-consistent electrostatics of Coulomb-interacting electrons) in a nearly constant area $S_\mu$.

Application of a front gate voltage $V_{FG}$ produces electric field which affects the confining potential and thus changes both the electron density distribution and $S_\mu$. The number of electrons in the island changes because both density and the area $S_\mu$ change. An increased $n$ is accompanied by the shift to a higher $B$ of the $i=1$ QH plateau (with Aharonov-Bohm oscillations superimposed). An increase in $S_\mu$ is observed as a smaller Aharonov-Bohm period $\Delta B$. Thus, the two effects are measured directly, independent of each other. From the dependence of the period $\Delta B$, and thus $S_\mu$, on $V_{FG}$ we obtain the electric-field-induced $\Delta S_\mu$, the electron orbit area change corresponding to addition of one electron inside the orbital at $\mu$. We also observe a systematic shift of the midpoint of the range of the Aharonov-Bohm oscillations $B_M$, which, assuming it corresponds to a fixed filling $\nu=1$, also yields the area $\Delta S_\mu = 2\pi \ell^2$ occupied by one electron. These two experimentally-independent ways to obtain $\Delta S_\mu$ agree very well. Additionally, we compare these experimental results to $\Delta S_\mu$ obtainable from the edge depletion models of Chklovskii et al.[13] and Gelfand and Halperin,[14] and find a reasonable agreement between $\Delta S_\mu$ obtained from these models and in the experiments.

## II. EXPERIMENTAL RESULTS

Our device is based on a very low disorder, high mobility modulation-doped AlGaAs/GaAs heterojunction,[15] which had a 2D electron system (2DES) with density $n_0 = 9.7 \times 10^{10}\,\text{cm}^{-2}$ (achieved after exposing the sample to red light at 4.2 K). Ohmic contacts were prepared on a pre-etched mesa. Next, a 2DES island of lithographic radius $R = 1300$ nm was defined by electron beam lithography (using proximity correction) and a self-aligned lift-off process (see Fig. 1). The 50 nm thick Au/Ti gate metal was deposited in chemically etched shallow trenches, 82 nm deep, reaching below the $\delta$-doping, while the 2DES is 215 nm below the surface. The four independent front gates are contacted separately.

All the measurements presented here were performed with the device in the 10.2 mK $^3$He-$^4$He bath in the tail of the mixing chamber of a top-loading into mixture dilution refrigerator. Extensive cold filtering in the electrical leads reduces the electromagnetic background incident on the sample to $5 \times 10^{-17}$ W.[16] We measure $R_{XX} = V_X / I_X$ as a function of magnetic field $B$ using a lock-in technique at 5.4 Hz. We typically use $I_X = 200$ pA rms in this paper, although reducing the current to 100 pA reveals moderate electron heating effects. Each $R_{XX}$ vs $B$ trace was measured at a fixed value of $V_{FG}$, defined as the average front gate voltage. A small differential front gate bias was applied in order to fine-tune the two constrictions for symmetry of tunneling amplitudes (to increase the amplitude of the oscillations).

In this paper we focus exclusively on the $i=1$ QH plateau in the island. The experimental results in this regime are summarized in Figs. 2, 3 and in the inset of Fig. 4. In general, we observe the Aharonov-Bohm oscillations in $R_{XX}$, superimposed on a smooth background magnetoresistance coming from the 2D bulk outside the island.[3] Fig. 2 presents a typical directly measured $R_{XX}$ vs. $B$ trace. The oscillations are clearly periodic with period $\Delta B$; for example, the trace shown in Fig. 2 ($V_{FG} \approx 0$) has $\Delta B = 1.87$ mT. Fig. 3 presents the oscillatory $\delta R_{XX}$ as a function of $B$ (that is, $R_{XX}$ with the smooth background subtracted), for several positive values of $V_{FG}$. The period $\Delta B$ for each of these traces decreases with increasing $V_{FG}$. It is evident that the magnetic field intervals where the Aharonov-Bohm oscillations occur shift to higher magnetic fields when $V_{FG}$ is increased. To quantify this behavior, we define $B_M$ as the midpoint of the magnetic field range in which the oscillations occur (for the given $V_{FG}$). For example, $B_M \approx 2.80$ T for the $V_{FG} = 150$ mV trace. The inset in Fig. 4 shows the dependence of thus determined $B_M$ on $V_{FG}$, which is approximately linear in the range of the voltages studied.

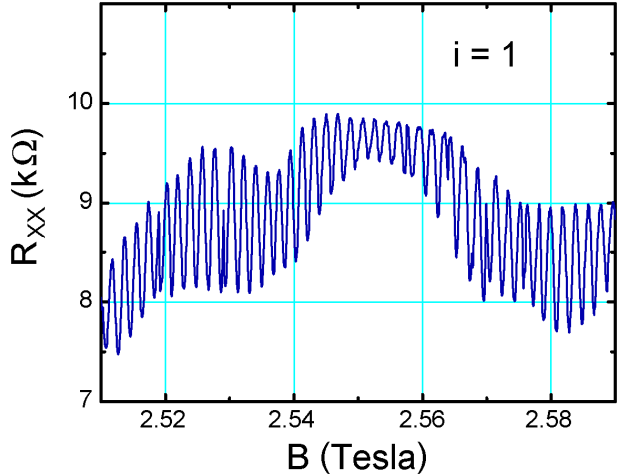

**FIG. 2.** Directly measured Aharonov-Bohm oscillations in $R_{XX}$ on $i = 1$ quantum Hall plateau. Front gate voltage $V_{FG} \approx 0$, bath temperature of 10.2 mK. Oscillations are superimposed on a smooth background due to conduction in the 2D electron "bulk", outside of the island.

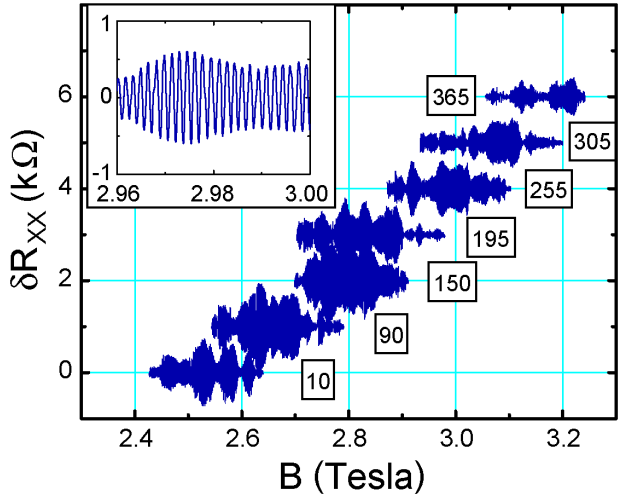

**FIG. 3.** Aharonov-Bohm oscillations $\delta R_{XX}$ as a function of $B$ for several values of positive front gate voltage $V_{FG}$, given in the labels next to each trace (in mV). All the traces occur on the $i = 1$ quantum Hall plateau, and have been displaced vertically in steps of 1 kΩ for clarity (1 kΩ corresponds to tunneling conductance 0.04 $e^2/h$). Each trace contains approximately 100 oscillations with a well defined period $\Delta B$, which depends on $V_{FG}$. Inset: a blow up of the $V_{FG} = 255$ mV trace shows the regularity of the oscillations.

### III. ANALYSIS AND DISCUSSION

#### A. The edge depletion models

The 2D electron island is defined by the depletion potential of the etch trenches. We use a model based on that of Gelfand and Halperin[14] (GH) to calculate the resulting electron density profile $n(r, V_{FG} = 0)$, assumed to be rotationally-symmetric. Briefly, this $B = 0$ classical electrostatics model includes effects of the surface charge on the side wall of the etched mesa due to the GaAs surface states, and the resulting ionized donors of the intentional 2D $\delta$-doping. GH obtain an analytic expression for the

- 4 -

density profile $n_{GH}(x) = F_{GH}(x/W)n_0$ for a linear edge, where $x$ is the coordinate normal to the edge ($x = -W$ at the lithographic edge), $W$ is the depletion length, $n_0$ is the 2D "bulk" electron density, and $F_{GH}$ is the function given in their Eq. (7). In a strong magnetic field $W \gg \ell$, the magnetic length, and this analytic expression agrees very well with the profile obtained in a Hartree-Fock calculation[14] for $x$ such that $n_{GH}(x)/n_0 > 0.3$. Since $W \ll R$, the island lithographic radius, we adapt the GH density profile to our circular geometry:

$$n_{GH}(r) = F_{GH}[(R-r)/W]F_{GH}[(R+r)/W]n_0, \qquad (1)$$

where $r$ is the radial distance from the center of the island.

The effect of the front gate bias is modeled following Chklovskii, Shklovskii, and Glazman[13] (CSG). They obtain an analytic expression for the electron density profile $n_{CSG}(x) = F_{CSG}(x/L)n_0$ for a linear edge ($x = -L$ at the gate edge), where $F_{CSG} = [(x+L)/(x-L)]^{1/2}$, and the length parameter $L = \varepsilon \varepsilon_0 V_{FG} / \pi e n_0$. Again, since $L \ll R$ for our $V_{FG}$ range, we adapt the CSG density profile to our circular geometry:

$$n_{CSG}(r) = F_{CSG}(r/L)n_0 = \left( \frac{(R+2L)^2 - r^2}{R^2 - r^2} \right)^{1/2} n_0. \qquad (2)$$

The main effect of a quantizing magnetic field is to open the QH gap at $\mu$, causing creation of "incompressible" and "compressible" regions.[13,14,17] The effect on electron transport properties is great, but the $B = 0$ electron density profile is not perturbed very much. This is because a variation of electron density produces large electrostatic charging energy, which must be compensated by the fraction of the QH gap energy gained per displaced electron. Recently, Hartree-Fock calculations for up to 300 electrons confined in an island for QH filling $i = 2$ have been reported for both parabolic and a bell-shaped confining potentials.[18] Certain qualitative similarities to the behavior in our samples are apparent, and perhaps future Hartree-Fock work can more closely model confinement in the electron interferometer samples.

**B. The Aharonov-Bohm quantization**

As mentioned in the Introduction, in this sample, in the integer QH regime, the Aharonov-Bohm oscillations arise from the modulation of quantum interference of electron paths by magnetic flux $\Phi$ enclosed by the two counterpropagating edge channels, coupled by tunneling (Fig. 1). The edge channels follow the equipotentials of the edge depletion potential, where the resulting electron density is such that Landau level filling $\nu = hn/eB \approx i = 1$ (note that $\nu \propto n$ in a given $B$). These approximately circular electron island states are quantized by the Aharonov-Bohm condition of one state per $\Phi_0$ per spin-polarized LL:

$$\Phi = BS_m = m\Phi_0, \qquad (3)$$

where the azimuthal quantum number $m = 0, 1, 2, \ldots$, $\Phi_0 = 2\pi\hbar/e = h/e$, and $S_m$ is the area enclosed by the $m$-th electron state. This quantization follows from equating the Aharonov-Bohm Berry phase of the $m$-th electron state to $2\pi m$, required for single-valued wave functions. In the low temperature, low excitation extreme quantum limit, the current is carried by electrons at the chemical potential $\mu$, and each oscillation in $R_{XX}$ signals the crossing of a single-electron quantized state with $\mu$. Consequently, the Aharonov-Bohm oscillation period $\Delta B$ corresponds to a flux change by one $\Phi_0$ through $S_\mu$ (the area enclosed by the electron state at $\mu$):

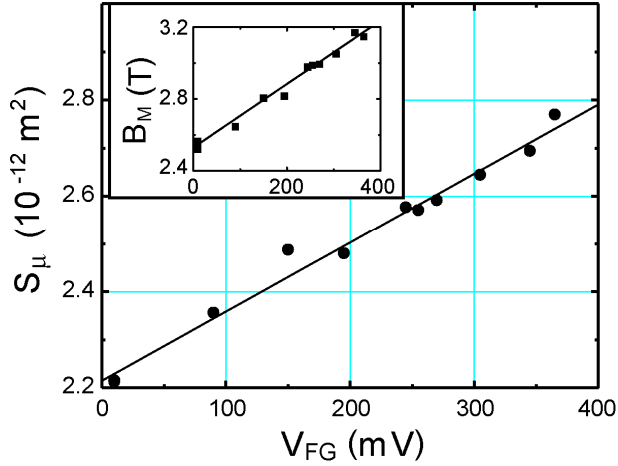

**FIG. 4.** Dependence of island area $S_\mu$ on $V_{FG}$. Each value of $S_\mu$ (circles) was determined from the Aharonov-Bohm period $\Delta B$. The dependence is approximately linear in the range of $V_{FG}$ studied; the solid line is a least squares fit to $S_\mu = a + bV_{FG}$, giving $a = 2.21\times10^{-12}$ m$^2$ and $b = 1.44\times10^{-12}$ m$^2$/V. Diameter $\sqrt{4S_\mu/\pi} \approx 1.8$ µm. Inset: the midpoint of the oscillations $B_M$ vs $V_{FG}$. A linear fit $B_M = c + dV_{FG}$ gives $c = 2.53$ T and $d = 1.77$ T/V.

$$S_\mu = \Phi_0/\Delta B. \tag{4}$$

Thus determined $S_\mu$ are plotted in Fig. 4 for several $V_{FG}$.

In a magnetic field sweep, at a fixed $V_{FG}$, as discussed in the Introduction, the density of the electron states in each LL is proportional to $B$. The electron density is pretty much constant, because disturbing the electron density results in a huge charging energy. For example, if each of the 100 observed oscillations resulted from transfer of one electron to the island, the total charging by $100e$ would result in ~10 eV charging energy; if an $i = 2$ ring containing 100 "extra" electrons were to form just within the $i = 1$ edge channel, thus allowing shrinkage of $S_\mu$, the charging energy would be ~2 eV, still enormous.

The chemical potential $\mu$ in the open island follows $\mu$ in the 2D bulk; LLs crossing $\mu$ results in de Haas-van Alphen and Shubnikov-de Haas oscillations. On a QH plateau, where the Aharonov-Bohm oscillations are observed, $\mu$ resides in the localized states between two LLs. The fixed number of $N$ electrons occupying the nearly constant area $S_\mu$ is accomplished by creating localized LL quasiparticles or quasiholes within the island edge ring. The Aharonov-Bohm oscillations result from a quantized single-electron state crossing $\mu$, which modulates the tunneling amplitude, accompanied by a microscopic electron population redistribution within $S_\mu$. Each oscillation corresponds to a change by one $\Phi_0$ in the flux through $S_\mu$, but at the higher field $B + \Delta B$ the area per $\Phi_0$ is less; thus the area $S_\mu$ remains nearly constant, and the number of electrons within $S_\mu$ is constant too. Since the quantization condition Eq. (3) is equivalent to the increment by $2\pi$ of the Berry phase of the electron wave function, we recover the Aharonov-Bohm periodicity of the constructive-destructive interference.

With increasing $V_{FG}$, more electrons are attracted to the 2DES island and the constrictions. Since tunneling amplitude is exponentially sensitive to the tunneling distance, the position of the tunneling links at the saddle points in the constrictions is nearly fixed, but the electron density $n_C$ at these positions increases with increasing $V_{FG}$. Accordingly, in order to remain on the $i = 1$ QH plateau, the applied $B$ must be increased. Within the island, the edge channels must follow the constant electron density contours with density equal that in the constrictions, $n_C$, and move outward, away from the island center. This edge channel density increase is confirmed by the shift to higher $B$ of the midpoint of the range of the Aharonov-Bohm oscillations: $n(r_\mu) \approx B_M/\Phi_0$ at QH filling $i = 1$, where $r_\mu$ is the radius of the electron orbit at $\mu$. The $B_M$ vs $V_{FG}$ dependence is shown in the inset of Fig. 4; the slope $dB_M/dV_{FG} = 1.77$ T/V corresponds to $dn(r_\mu)/dV_{FG} = 4.3\times10^{14}$ 1/m$^2$V. Similarly, the area $S_\mu$ is expected to increase with increasing $V_{FG}$ because of reduced mesa depletion. An approximately linear dependence of $S_\mu$ on $V_{FG}$ is indeed obtained from the Aharonov-Bohm period $\Delta B$, see Fig. 4, the slope $dS_\mu / dV_{FG} = 1.44 \times 10^{-12}$ m²/V.

Reference 5 reported and analyzed Aharonov-Bohm oscillations observed in a Coulomb Island of lithographic radius $R = 750$ nm, separated from the 2D bulk by two narrow (300 nm wide) point contacts, showing conductance steps at $B = 0$. The gates were deposited on GaAs surface, with no etch trenches; a fixed negative $V_{FG}$ was applied to produce the confining potential. The analysis of Ref. 5 invoked a large negative correction to the Aharonov-Bohm period, Eq. (4), which was justified by a seemingly strong dependence of the product $i\Delta B$ on the island QH filling $i$, assumed equal to that in the bulk. Actually, from the data in Figs. 2 and 3 and the text of Ref. 5, it is clear that the island edge ring filling was ~1.7 times smaller than in the bulk. Accordingly, the Aharonov-Bohm oscillations regions centered on $B = $ 5.1, 2.6, 1.85 and 1.4 T should be identified as belonging to the $i = $ 1, 2, 3 and 4 QH plateaus, respectively. Using this filling factor assignment, the corresponding periods $\Delta B = $ 5.3, 2.7, 2.0 and 1.3 mT scale well, so that the product $i\Delta B \approx 5.4$ mT is constant (within the experimental uncertainty). This yields the area $S_\mu = \Phi_0 / i\Delta B = 7.6 \times 10^{-13}$ m² and the radius $r_\mu = 495$ nm, values independent of $i$ and reasonable. The $i = $ 1, 2 and 4 Aharonov-Bohm oscillations reported[19] for the device of Ref. 3 also yield a constant product $i\Delta B \approx 2.7 \pm 0.1$ mT. This analysis involves no correction to our Eq. (4) due to the confining potential.

**C. The front gate voltage period**

Although not done in experiments reported here, in principle it is possible to sweep front gate voltage $V_{FG}$ continuously (at a fixed $B$). We denote $\Delta V_{FG}$ the expected period, which induces a change $\Delta N = 1$ in the number of electrons $N$ within $S_\mu$. Application of $\Delta V_{FG}$ results in a change $\Delta S_\mu$, which can be linearized as $\Delta S_\mu = (dS_\mu / dV_{FG})\Delta V_{FG}$ for large $N \gg 1$ (in this device $N \approx 2000$). Since the area $\Delta S_\mu$ can be occupied by precisely one electron per spin-polarized Landau level, $\Delta S_\mu$ can be identified with the area between consecutive quantized electron orbits $S_{m+1} - S_m$ in the vicinity of $\mu$. Hence, from the Aharonov-Bohm quantization condition Eq. (3), substituting $B_M$ for $B$:

$$\Delta S_\mu = \Phi_0 / B_M . \qquad (5)$$

Note that Eq. (5) simply gives $\Delta S_\mu = 2\pi\ell^2$ with magnetic length corresponding to $B_M$. Combining these two expressions for $\Delta S_\mu$, we get an expression for $\Delta V_{FG}$ in terms of experimentally obtained quantities:

$$\Delta V_{FG} = \Phi_0 / (dS_\mu / dV_{FG}) B_M . \qquad (6)$$

We calculated $\Delta V_{FG}$ using the experimental value of $dS_\mu / dV_{FG}$ and the values of $B_M$ (see Fig. 4). We find that $\Delta V_{FG}$ decreases with increasing $V_{FG}$, but the product $S_\mu \Delta V_{FG}$ is approximately constant, independent of $V_{FG}$ (within the experimental uncertainty of 2%), see Fig. 5(a). The product $S_\mu \Delta V_{FG}$ is proportional to the inverse integrated compressibility of the 2D electrons in the island: $dn/d\mu \propto dn/dV_{FG} \propto (dN/dV_{FG})/S_\mu = \Delta N / S_\mu \Delta V_{FG} = 1/S_\mu \Delta V_{FG}$. In this perspective, the constancy of $S_\mu \Delta V_{FG}$ is very fundamental: it stems from the fixed density of states in the island area encircled by the $i=1$ edge channel, that is, in one spin-split Landau level. The Aharonov-Bohm signal originates in the interference of electrons moving in the current-carrying edge channel. Under increasing positive $V_{FG}$, the edge channel radius increases, but, so long as the island remains on the $i=1$ plateau, the opening of the QH gap precludes population of the next Landau level in the interior of the island (recall that $B$ is fixed). That is, the population of the island by additional electrons occurs by an enlargement of the edge channel radius, as opposed to the $B=0$ case, where additional electrons are induced throughout the area of the island.

The constancy of the product $S_\mu \Delta V_{FG}$ can also be viewed as a constant differential capacitance per unit area, $C/S_\mu = 64$ µF/m², where $C = dQ/dV = e/\Delta V_{FG}$ is the differential capacitance between the front gates and the 2D electron island. The effect must be quantum, since the classical island capacitance can be estimated to change by a factor of 1.5, from the data of Fig. 4. A constant differential capacitance in QH regime has been reported in quantum antidots[20] and in mesoscopic Si MOSFETs.[21]

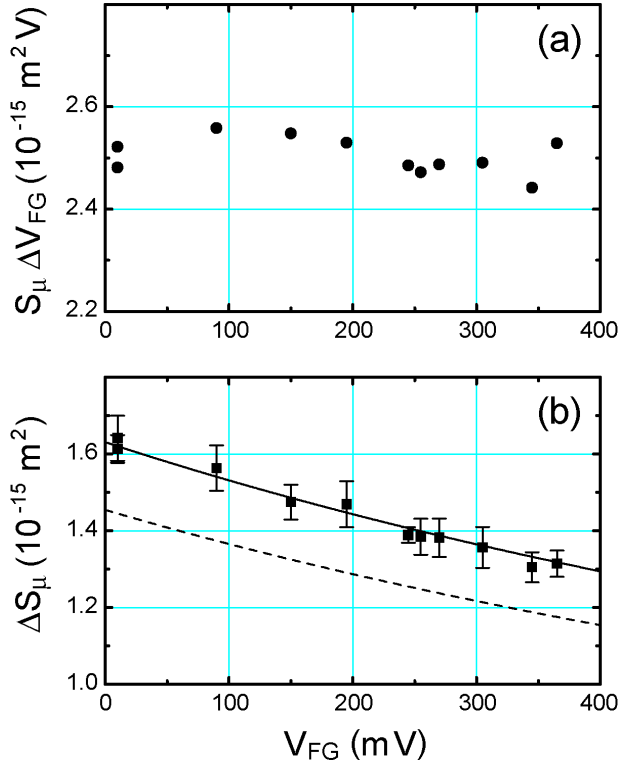

**FIG. 5.** (a) Dependence of the product $S_\mu \Delta V_{FG}$ on $V_{FG}$, where the increment $\Delta V_{FG}$ induces one more electron within the area $S_\mu$ at a fixed $B$. The product $S_\mu \Delta V_{FG}$ is constant within $\pm 2\%$ in the range of $V_{FG}$ from 0 to 400 mV, while $S_\mu$ increases by 27% in the same range (Fig. 4). (b) One electron state area $\Delta S_\mu$ vs $V_{FG}$. The squares with error bars ($\Delta S_\mu$) are determined from the experiment as $\Phi_0/B_M = 2\pi\ell^2$, Eq. (5), while the solid line ($\Delta S_\mu^*$) is obtained from the fit of the experimental Aharonov-Bohm periods $\Delta B$, Eq. (7). The dashed line gives $\Delta S_\mu$ calculated by integrating the edge depletion model electron density profile with no adjustable parameters. The 13% discrepancy between the model calculation and the experiment is surprisingly small, considering the several approximations made and the idealized, simplified geometry of the model.

Using the experimental fact that the product $S_\mu \Delta V_{FG}$ does not depend on $V_{FG}$, and the linearization $\Delta S_\mu = (dS_\mu / dV_{FG}) \Delta V_{FG}$, we obtain:

$$\Delta S_\mu^*(V_{FG}) = \left.\frac{dS_\mu}{dV_{FG}}\right|_{V_{FG}} \frac{S_\mu(0)\Delta V_{FG}(0)}{S_\mu(V_{FG})}, \qquad (7)$$

where $S_\mu(0)$ and $\Delta V_{FG}(0)$ are evaluated at $V_{FG}=0$. The asterisk in $\Delta S_\mu^*$ serves to distinguish it from $\Delta S_\mu$ calculated using Eq. (5). Equation (7) is of interest because it can be used to obtain $\Delta S_\mu^*$ from entirely different experimental input, the dependence of the Aharonov-Bohm period $\Delta B$ on $V_{FG}$, than those used in Eq. (5), that is, the dependence of $B_M$ (positions of QH filling $i=1$ in the island) on $V_{FG}$.

The results obtained from these two routes are given in Fig. 5(b), where $\Delta S_\mu^*$ (solid line) is calculated using the slope $dS_\mu / dV_{FG}$ of the linear dependence of $S_\mu$ on $V_{FG}$ (Fig. 4) and $\Delta V_{FG}(0) = 1.13$ mV, Eq. (6). The error bars in $\Delta S_\mu$ are most likely overestimated by taking an error in $B_M$ equal to half the range of the magnetic field at which the oscillations occur. As can be seen, the agreement between $\Delta S_\mu$ and $\Delta S_\mu^*$ is very good, well within the error bars. This agreement confirms that the Aharonov-Bohm quantization condition, Eqs. (3) and (4), describes well the basic physics of the electron interferometer devices in the QH regime studied in this work and in Ref. 3.

### D. Comparison to the edge depletion models

As stated in Section A., the primary 2D electron confining potential is created by the depletion potential of the etch trenches. The resulting 2D electron density profile is affected by the electric field of the front gates. A relatively robust way to evaluate the effect of $V_{FG}$ on the electron island is to calculate $\Delta S_\mu$ due to $\Delta V_{FG}$, the voltage increment needed to attract one more electron to $S_\mu$. To this end, we evaluate $n_{GH}(r)$ from Eq. (1), using the known heterostructure and lithographic parameters and dimensions as input. We then calculate $n(r, V_{FG}) = n_{GH}(r) F_{CSG}(r, V_{FG})$, using $F_{CSG}$ from Eq. (2), where the $V_{FG}$-dependence is contained in $L \propto V_{FG}$ (for the present device, $L/V_{FG} = 0.23$ nm/mV). Such a calculation is not self-consistent, but, in absence of a self-consistent result, is justified by: (i) in this paper, front gate voltage in the range of $0 < V_{FG} < 400$ mV is only a perturbation to the larger depletion effect of the etch trenches, and (ii) we use only $n(r, V_{FG})$ integrated over the area of $S_\mu$, which excludes the low density tail of $n(r, V_{FG})$, and thus should reduce relative error due to non-self-consistency of the calculation, because such error is larger for nearly depleted regions of small electron density.

Integrating $n(r, V_{FG})$ over the experimental Aharonov-Bohm area $S_\mu$, we determine $\Delta V_{FG}$ such that the number of electrons within $S_\mu$ increments by one: $\Delta N = \int n(V_{FG} + \Delta V_{FG}) dS - \int n(V_{FG}) dS = 1$. In particular, we obtain $\Delta V_{FG}(0) = 1.01$ mV starting at $V_{FG} = 0$. Using Eq. (7), the increment $\Delta V_{FG}(0)$ gives the one electron state area $\Delta S_\mu$, shown by the dashed line in Fig. 5(b). The model calculation described above involves no adjustable parameters, has experimental input via $S_\mu$, and uses an idealized device geometry. The surprisingly small difference (13%) between the experimental and the calculated $\Delta S_\mu$ indicates that the CSG model adequately describes the effect of front gates on electron density profile for moderate $V_{FG}$.

### IV. CONCLUSIONS

We reported electron quantum interference experiments on the $i = 1$ quantum Hall plateau performed with an electron interferometer device. When tunneling between edge channels occurs, in the quantum-coherent regime, Aharonov-Bohm oscillations with period $\Delta B$ were observed as a function of magnetic field $B$. In this regime, the quantized electron states encircling the island are quantized by the Aharonov-Bohm condition: the magnetic flux through the area of the closed electron path at the chemical potential $\mu$ satisfies $\Delta \Phi = \Delta B \, S_\mu = \Phi_0$, where $\Phi_0 = h/e$. Each oscillation corresponds to a change by one $\Phi_0$ in the flux through $S_\mu$. Microscopically, each Landau level has one electron state per $\Phi_0$, and addition of flux $\Phi_0$ leads to the outermost empty state crossing $\mu$, that is, becoming occupied.

Simultaneously, addition of $\Phi_0$ to the interior of the island creates a hole state in the otherwise occupied LL. Thus the area $S_\mu$ remains nearly invariable, and the number of electrons within $S_\mu$ is constant, too. We experimentally determined the dependence of $S_\mu$ on the front gate voltage, and conclude that the Aharonov-Bohm quantization condition does not require significant corrections due to the confining potential. These results can be interpreted as a constant integrated compressibility of the island (on a quantum Hall plateau) with respect to the front gates. We also analyzed experimental results using two classical electrostatics models: one modeling the 2D electron density due to depletion from an etch trench, and another modeling the gate voltage dependence of the electron density profile in the island. We conclude that the models adequately describe the effect of front gates on electron density profile for moderate biases.

**ACKNOWLEDGEMENTS:** This work was supported in part by the National Science Foundation under Grant DMR-0303705 and by NSA and ARDA through US Army Research Office under Grant DAAD19-03-1-0126.

**References:**

1. V. J. Goldman and B. Su, Science **267**, 1010 (1995); V. J. Goldman, I. Karakurt, J. Liu, and A. Zaslavsky, Phys. Rev. B **64**, 085319 (2001).

2. V. J. Goldman, J. Korean Phys. Soc. **39**, 512 (2001).

3. F. E. Camino, W. Zhou, and V. J. Goldman, Phys. Rev. B **72**, 075342 (2005).

4. P. L. McEuen *et al*., Phys. Rev. Lett. **66**, 1926 (1991).

5. B. J. van Wees *et al*., Phys. Rev. Lett. **62**, 2523 (1989); L. P. Kouwenhoven *et al*., Surf. Science **229**, 290 (1990).

6. B. I. Halperin, Phys. Rev. B **25**, 2185 (1982); A. H. MacDonald and P. Streda, Phys. Rev. B **29**, 1616 (1984).

7. X. G. Wen, J. Mod. Phys. B **6**, 1711 (1992).

8. P. L. McEuen *et al.*, Phys. Rev. Lett. **64**, 2062 (1990).

9. J. K. Wang and V. J. Goldman, Phys. Rev. Lett. **67**, 749 (1991); Phys. Rev. B **45**, 13479 (1992).

10. See, e.g., S. M. Girvin, *The Quantum Hall Effect*, Les Houches Lecture Notes (Springer-Verlag, New York, 1998).

11. I. Karakurt, V. J. Goldman, J. Liu, and A. Zaslavsky, Phys. Rev. Lett. **87**, 146801 (2001).

12. The exact integer filling $i$ is the principal quantum number of the QH state, defined as the quantized value of the Hall conductance $\sigma_{XY}$ in units of $e^2/h$.

13. D. B. Chklovskii, B. I. Shklovskii, and L. I. Glazman, Phys. Rev. B **46**, 4026 (1992).

14. B. Y. Gelfand and B. I. Halperin, Phys. Rev B **49**, 1862 (1994).

15. Heterojunction material of the device in this paper is the same as in Ref. 9. The sample studied in this paper is different from that studied in Ref. 3.

16. I. J. Maasilta and V. J. Goldman, Phys. Rev. B **55**, 4081 (1997).

17. C. de C. Chamon and X. G. Wen, Phys. Rev B **49**, 8227 (1994).

18. N. Y. Hwang, S. R. E. Yang, H. S. Sim, and H. Yi, Phys. Rev. B **70**, 085322 (2004).

19. W. Zhou, F. E. Camino, and V. J. Goldman, to appear in Proc. 24th Intl. Conf. Low Temp. Physics (2005).

20. I. J. Maasilta and V. J. Goldman, Phys. Rev. B **57**, R4273 (1998).

21. D. H. Cobden, C. H. W. Barnes, and C. J. B. Ford, Phys. Rev. Lett. **82**, 4695 (1999).